\documentclass[journal]{IEEEtran}
\usepackage[utf8]{inputenc} 
\usepackage{hyperref}       
\usepackage{url}            
\usepackage{booktabs}       
\usepackage{amsfonts}       
\usepackage{nicefrac}       
\usepackage{microtype}      
\usepackage{graphicx}
\usepackage{enumerate}
\usepackage[dvipsnames]{xcolor}
\usepackage{titlesec}
\usepackage[noadjust]{cite}
\usepackage[normalem]{ulem}
\usepackage[colorinlistoftodos]{todonotes}
\usepackage{longtable}

\ifCLASSINFOpdf
\else
\fi
\usepackage{amsmath}
\begin{document}
%
\title{Power Modeling for Effective Datacenter Planning and Compute Management}
%
%
%

\author{Ana Radovanovic, 
        Bokan Chen, 
        Saurav Talukdar, 
        Binz Roy, 
        Alexandre Duarte,
        and Mahya Shahbazi
\thanks{The authors are with Google, Inc. Mountain View, CA, 94043 (Email: anaradovanovic@google.com, bokanchen@google.com, stalukdar@google.com, binzroy@google.com, alexandredu@google.com, mahyash@google.com)
.}
}

\maketitle

\begin{abstract}
Over the past decades, there has been a global growth in datacenter capacity, power consumption and the associated costs. Thus, an accurate mapping of the datacenter's compute resources (CPU, RAM, etc.) and hardware types (servers, accelerators, etc.) to power consumption has emerged as an essential requirement for major Web and cloud service providers. Therefore, this paper presents two types of statistical power models designed and validated for accuracy, simplicity, interpretability, and applicability to all hardware configurations and workloads across hyperscale datacenters of Google fleet. The models have been designed (and already deployed in production) based on: 1) individual machine level data, and 2) the entire MW-scale datacenter Power Distribution Unit (PDU) level. In addition, they are developed with several use cases in mind, including: cost- and carbon-aware load management, power provisioning, and infrastructure rightsizing. 
To the best of our knowledge, this is the largest scale datacenter power modeling effort using real operational data, in both the scope of diverse datacenter planning and real-time management use cases, as well as the variety of hardware configurations and workload types used for modeling and validation. We demonstrate that the deployed statistical modeling techniques, while simple and scalable, 
predict power with less than 5\% Mean Absolute Percent Error (MAPE) for more than 95\% diverse Power Distribution Units across Google fleet. This performance matches the reported accuracy of the previous state-of-the-art methods, while using significantly less features and covering a wider range of use cases.

\end{abstract}


%
\IEEEpeerreviewmaketitle
\section*{Nomenclature}
\noindent
\begin{tabular}{l p{7cm}}
$r$ & The index indicating PDU operation regimes. $r\in \{low, medium, high\}$.\\ 
$d$ & The index indicating day. \\
$\mathcal{T}$ & The set of $288$ 5-minute time periods within a day. $t\in\mathcal{T}$. \\ 
$\mathcal{C}$ & The set of PDUs in a cluster. $PDU\in\mathcal{C}$ \\
$cpu_-^{PDU}$ & Minimum PDU level CPU usage. \\ 
$cpu_+^{PDU}$ & Maximum PDU level CPU usage. \\
$\hat{\large{\cdot}}$ & The predicted value of a variable. \\
$\bar{\large{\cdot}}$ & The average value of a variable. \\
$u^{PDU}_{CPU}$ & PDU level CPU usage.  \\
$\lambda$ & Segment length in the per-PDU model. \\
$\alpha^r$ & Intercept of the per-PDU model in operation regime $r$. \\ 
$\beta^r$ & Slope of the per-PDU model in operation regime $r$. \\
$u^{PDU}_{POW}$ & PDU power consumption. \\
\end{tabular}
\\

\noindent
\begin{tabular}{l p{7cm}}
$u_{CPU}^m$ & CPU usage of individual machines. \\
$u_{POW}^m$ & Machine level power consumption. \\
$C^m$ & Machine configuration. \\
$P_-^m$ & Idle power of a machine. \\
$P_+^m$ & Maximum power usage of a machine. \\
$l^d$ & Dedicated label of a machine.\\
$u_{POW}^{IT}$ & IT power usage on a PDU. \\
$u_{POW}^{O}$ & The overhead power on a PDU. \\
$N_{F_i}$ & Platform count of each platform family.\\
$u_{CPU}^{F_i}$ & CPU usage per platform family. \\
$P^{N}$ & Maximum networking power. \\
$P^{C}$ & Maximum cooling power. \\
$A^{P}$ & Power architecture. \\
$P_-^{PDU}$ & Total idle power of a PDU. \\
$P_+^{PDU}$ & Maximum IT power of a PDU. \\
\end{tabular}
\section{Introduction}\label{sect:intro}
Datacenters are warehouse-sized computing systems that operate around the clock to support large-scale Internet services worldwide, while enabling fast growth of the IT industry and transformation of the economy. The demand for computing resources and datacenter power worldwide has been continuously growing, contributing to approximately 1\% of the total electricity usage \cite{masanet2020recalibrating}. Hyperscale datacenters are known to consume tens of thousands megawatt hours of electricity per day (for comparison, the entire city of San Francisco consumes ~20000MWh per day \cite{cacommission}). Furthermore, new datacenter machine upgrades are planned based on the peak power, and the overall infrastructure costs are highly dependent on resource (e.g. CPU) utilization \cite{fan2007power, choi2008profiling}. Given the rapid growth of datacenter workload globally (more than sixfold in a decade) \cite{masanet2020recalibrating}, novel methodologies for improving datacenter power and energy efficiency are essential due to the high economic, environmental and performance impacts \cite{dayarathna2015data}.

Universally accurate power models enable more efficient datacenter planning and compute load management. 
Therefore, in this paper, two types of accurate and light-weight power models are developed to characterize power consumption of a datacenter based on workload CPU usage. The models are designed with the following use cases in mind: 1) Real-time estimation of datacenter power consumption and its electricity-based carbon footprint (\cite{energy2019data, greenieee}); 2) Near-term (intraday and day-ahead) cost- and carbon-aware workload management (including software-controlled power capping \cite{fan2007power} and grid level demand response \cite{LiuChenBashWierman2012, liu2015datacenter, radovanovic2020our, energy2019data}); and 3) Power provisioning and rightsizing of future machine upgrades given monthly, quarterly, or yearly planned resource usage trends \cite{fan2007power}.

Even though power consumption in hyperscale datacenters is mainly driven by IT equipment \cite{masanet2020recalibrating}, quantifying the impact of the above-mentioned use cases on the overall datacenter campus power requires understanding of its relationship to the lower level, typically metered, power domains called Power Distribution Units (PDUs), which encapsulate all servers, networking equipment and storage. The goal of this paper is to accurately relate PDU's workload resource usage to its power consumption at 5-minute time granularity. We develop two types of light-weight statistical power models using the vast amount of data across all Google's datacenter PDUs\footnote{Incorporating all PDUs across more than 20 Google datacenters globally, with more than 10 terawatt-hour yearly energy consumption \cite{googledatacenters,googleenergyconsumed}} with varying hardware configurations, machine platforms and workload types. 

The first type of the models, called \textit{Per-PDU} model, is piecewise-linear in CPU usage and is retrained daily for each PDU. The second involves two models, referred to as \textit{Unified}, and trained to: 1) predict a single machine's power, which is then aggregated to estimate PDU power consumption (Unified \textit{Machine} Model), and 2) directly predict PDU power (Unified \textit{PDU} model). The Unified power models utilize a machine leaning framework based on the Random Forest regression to capture the nonlinear dependence between machine/PDU power consumption, its CPU usage, machine/PDU's hardware characteristics, and other properties. These models are trained to capture any machine/PDU's configuration type deployed within Google's datacenter fleet. Note that all other machine- and PDU-related hardware and workload properties, including effects of task scheduling, CPU and voltage frequency scaling are implicitly captured by the proposed models.

The proposed models are trained using a modeling pipeline, running daily in an automated and parallelized manner, capturing data from all power domains across Google's datacenter fleet. While the proposed models ensure high and mutually comparable accuracy when predicting PDU power consumption in steady state, each model shows advantages in addressing specific use cases. High accuracy of the Per-PDU model across all CPU/power operating states and its piecewise linearity in CPU usage make it an ideal candidate for evaluating the impact of changes in resource usage on power consumption. Such changes in resource usage can happen when the flexible workload is shifted across time and space. On the other hand, the Unified models are best suited for quantifying effects of: 1) more significant compute load migrations across compute clusters, 2) perturbations caused by maintenance and demand response events, and 3) machine upgrades. Rightsizing of future datacenters is also another application of the proposed models.

The models have been deployed and rigorously validated for all power domains across Google's datacenters. Their high accuracy is demonstrated irrespective of the heterogeneity in different machine configurations (e.g., CPU-based general purpose server platforms, accelerator or GPU-based platforms), workload types (e.g., mail, video processing and streaming, distributed computing), platform  families (e.g., Intel  Cascade  Lake, IBM Power8, GPU based servers), platform level control mechanisms (e.g., Turbo Boost, DVFS), and power architectures.
In addition, the prediction error is characterized accurately in scenarios where the measured PDU power is larger than its prediction (i.e., for instances with underprediction error). This is a critical requirement for risk-aware control of PDU peak power usage, where the goal is to protect breaker limits or avoid exceeding a given power usage threshold (commonly referred to as power capping limit).


The rest of the paper is organized as follows: Section \ref{sect:prior_art} reviews the state of the art and presents the contributions of the current paper. Section \ref{sect:infra_modeling} describes the overall architecture of a datacenter, including the available power telemetry systems necessary for model verification as well as the training pipeline for the power models, along with a discussion on feature selection, model structure and complexity. Performance of the proposed models and their applicability to different use cases are discussed in Section \ref{sect:perf_eval}. Finally, Section \ref{sect:conclusion} concludes the paper.

\section{Prior art and contributions of the paper}\label{sect:prior_art}
There exist a body of literature focused on provisioning power consumption at a component, circuit, server, PDU and all the way to datacenter level \cite{dayarathna2015data}. While component level power models (e.g. processor) are typically designed to capture relationships between control signals and component states at different time scales \cite{dayarathna2015data, choi2008profiling, chen2005managing, liu2016fastcap}, they are usually designed for specific types of hardware and are hard to use for power provisioning in hyperscale datacenters that are highly heterogeneous in workload types, platform families, and platform level control mechanisms \cite{dev2020autonomous}.

On the other hand, statistical models (or sometimes referred to as software-based models) \cite{dayarathna2015data,10.1145/3390605} have proven effective in modeling either individual subsystems of a server such as CPU, memory, disk and network, or a virtual machine, server, server cluster, or a datacenter as a whole. The previous studies that mainly focused on power modeling of a single machine or a group of machines have been limited in the number of machine configurations, workload types, and discussed use cases (\cite{10.1145/3390605} and references therein). Also, modeling of power consumption at a datacenter level has mainly been addressed via stylized models \cite{GANDHI20101123, 10.1145/2492101.1555368, 10.1145/1592568.1592584} with insufficient performance validation.

While this paper shares some insights from the previously published work, its main contributions are summarized below:
\begin{itemize}
\item \textbf{Use-case-driven design:} The approach for model design and feature selection is strongly impacted by a wide range of targeted use cases, oriented toward more efficient, carbon-aware and dynamic workload management, as well as longer-term planning and rightsizing of future datacenters.
\item \textbf{Largest scale validation:} To the best of our knowledge, this is the largest scope system deployed for power modeling and validation, incorporating all power domains across Google's datacenter fleet with heterogeneous hardware configurations, workload types, and resource utilization levels. It is demonstrated that the generality and effectiveness in provisioning PDU power is achievable using only basic hardware and resource usage characteristics (such as CPU usage or CPU utilization, i.e., CPU usage divided by the total CPU capacity). It should be mentioned that the models discussed in this paper have been deployed and used for more than a year.
\item \textbf{High accuracy:} While the proposed models use significantly fewer features, they are applicable throughout the dynamic range of power domain's utilization, and their performance is well aligned with the reported accuracy of the related state-of-the-art methods \cite{fan2007power, rivoire2008comparison, davis2011no} 
\end{itemize}

\section{Datacenter power infrastructure layout and modeling}\label{sect:infra_modeling}

In this section, we describe a typical Google datacenter power system architecture, data collection pipeline, and the power modeling methodology including training and evaluation processes. We also provide a brief discussion on model complexity and scalability.

\subsection{Datacenter power architectures}\label{subsect:power_archit}
\begin{figure}
  \centering
  \includegraphics[width=\linewidth]{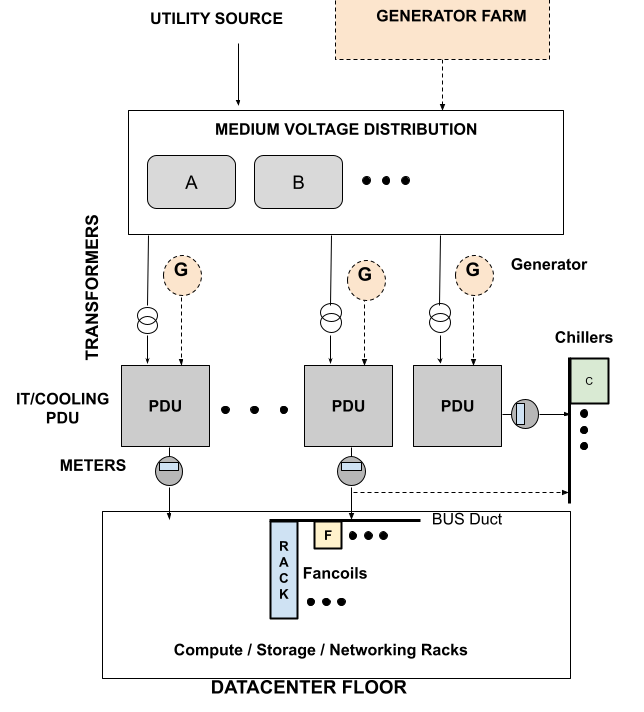}
  \caption{A simplified representation of datacenter power distribution hierarchy.}
  \label{power_arch}
\end{figure}

Figure \ref{power_arch} shows a simplified view of power architecture of a typical Google datacenter \cite{barroso2018datacenter}. Every datacenter is connected to the electricity grid via several medium-voltage feeders. Each medium-voltage distribution line is transformed to supply low-voltage PDUs. PDUs are typically connected to multiple BUS ducts. The BUS ducts supply power to the IT equipment, fan coils used for cooling) on the datacenter floor, and, in some cases, off-floor cooling equipment (e.g., chillers).

The IT equipment on the datacenter floor comprises compute, storage, and networking racks. A single PDU typically has a few thousand machines, and a handful of PDUs comprise a cluster. The PDUs in each cluster belong to a single job-scheduling domain, i.e., a common real-time scheduler that assigns computing tasks to its feasible machines. Generator backup is available to keep the datacenter running in the event of a grid power outage. Depending on the architecture, the PDUs in a cluster are either connected to a separate backup generators, or to a common backup generator that supports the medium-voltage line.

All the PDUs are metered and provide power measurements, which are used to train and validate the models discussed in the subsequent sections. Note that the power usage in a datacenter's campus can be accurately expressed as the sum of its individual PDU measurements inflated by a few percent overhead to account for unmetered  auxiliary loads such as office HVAC.


\subsection{The power modeling pipeline} \label{sect:power_model}
\begin{figure}[!b]
    \centering
    \includegraphics[width=\linewidth]{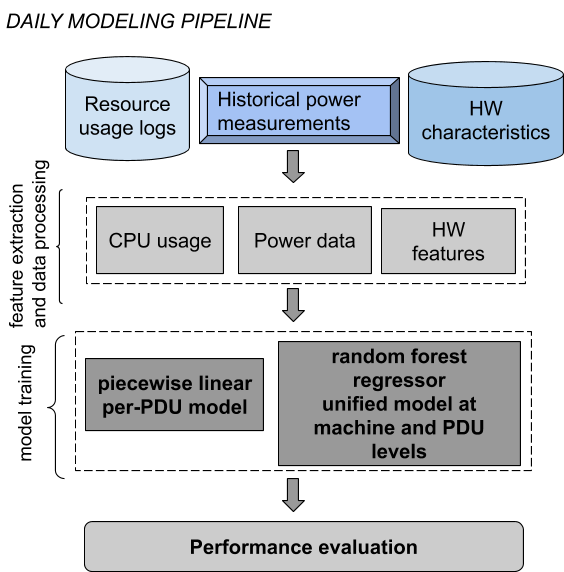}
    \caption{Data processing and training pipeline.}
    \label{data_pipeline}
\end{figure}

A power modeling pipeline is used for data extraction, processing, and training of two types of models to predict power consumption across Google's datacenter fleet at 5-minute granularity. 
The pipeline, shown in Figure \ref{data_pipeline}, consists of several components that: 
\begin{enumerate}
\item read historical resource and power usage data at 5-minute granularity, as well as hardware characteristics of each machine/PDU,
\item process raw data to detect and remove outliers, 
\item train models in a highly parallelized manner, and 
\item evaluate performance of the models with rigorous validation procedures adapted to match the targeted use cases.
\end{enumerate}

The measurements used by the pipeline includes: 1) real-time power consumption and CPU usage of every single machine across the fleet, collected through built-in power sensors and reported by the machine OS. The machine level power and CPU usage data is being stored in the resource usage and power logs and maintained as a Google internal database; 2) PDU power, which is collected through the power meters on each PDU and is stored in a separate Google internal database. In addition to these measurements, the data set also includes machine configuration information, as well as maximum and idle power of each machine, as discussed in more details in Section \ref{subsubsubsection:UMM}.

A data perprocessing step is applied before training the models to automatically detect and remove outliers from the data set. These outliers mainly correspond to atypical regimes in resource usage and power measurements caused by rare maintenance events, anomalies, losses in stored data, etc. For example, rare and short losses in the stored usage data are handled using linear interpolation of the time series. Other heuristics applied to automatically detect and remove typical outliers include:
\begin{itemize}
\item Smoothing using Exponentially Weighted Moving Average (EWMA) in case of a significant (and hardly possible) change in resource usage, e.g., more than $30$\% from one 5-minute time interval to the next;
\item Removing data instances for which changes in power measurements cannot physically match the corresponding change in resource usage from one 5-minute interval to the next. For example, data instances with $\Delta |\text{power usage}| > 20 \Delta |\text{resource usage}|$ for any two consecutive 5-minute measurements are excluded. This type of anomaly is most likely caused by erroneous power meter readings;
\item Filtering data instances with PDU's power measurements smaller than $80$\% of its daily median. It is observed that these power drops are extremely rare and coincide with 1-2 hour-long maintenance events.
\end{itemize}

The models are then trained on the filtered data set after removing the outliers. The model training and validation pipeline uses the actual power measurements collected from each PDU (from AC power meters) and machine (from DC machine power supply) across the datacenter fleet.

\subsection{Power Models}

As discussed above, PDUs are the fundamental power domain that contain servers, cooling and networking equipment. The telemetry system available in each PDU enables collection of power consumption data at 5-minute granularity, and, thereby, a supervised learning framework for modeling PDU power consumption.

While a datacenter infrastructure planning and real-time workload management are typically driven by trends in workload resource usage (CPU, RAM, disk usage), the analysis in this paper demonstrates that PDU power consumption can be accurately estimated using only its CPU usage as a resource usage feature. The general conclusion holds irrespective of the diversity in machine types, e.g., compute, storage, accelerators (TPU or GPU), etc. This might appear surprising for storage and accelerator machines where CPU usage is not the predominant driver of the power usage. However, storage machines are well known to have a narrow dynamic range of power \cite{dayarathna2015data}. In case of accelerators with weak CPU-power correlation, we also observed a narrow dynamic range. This narrow dynamic range results in a smaller impact on the power fluctuations when aggregated at PDU level. As it will be discussed in Section \ref{sect:perf_eval}, our proposed models are able to estimate the aggregate power at PDU level accurately irrespective of the PDU's diversity in machine types and workload. The accuracy of the models are demonstrated using the operational data across all PDUs in Google's datacenters.

The following subsections discuss the two types of power models. The first is the Per-PDU model, used for power provisioning and efficiency-aware workload management, such as load shifting in time and space, demand response, etc. The second type of models includes the so-called Unified Machine and PDU models, which are developed for optimized long-term planning such as power oversubscription and infrastructure rightsizing\cite{49032}.

\subsubsection{Per-PDU power model}
\label{ssec:perpdu}

To enable power efficiency and carbon-aware workload management, empowered by computationally tractable and scalable optimization formulations \cite{radovanovic2020our}, an accurate, light-weight, piecewise linear model is developed and deployed. A piecewise linear model structure enables easier and tractable integration of power information in the optimization objective and constraints of datacenter load management optimization problems.

\begin{figure}[h]
\centering
    \includegraphics[width=0.7\linewidth]{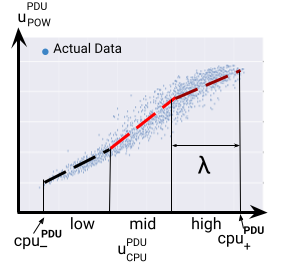}
    \caption{Structure of the piecewise linear model.}
    \label{fig:per_scatter}
\end{figure}

Extensive analysis of average 5-minute PDU power consumption $u_{POW}^{PDU}$, as a function of its average CPU usage $u_{CPU}^{PDU}$, indicates three distinct utilization regimes (Figure \ref{fig:per_scatter}). Furthermore, the higher a PDU's CPU usage, the higher its overall power consumption. Based on this observation, the linear models are trained for \textit{low}, \textit{medium} and \textit{high} CPU usage regimes, as defined below:
\begin{align}
    &\begin{cases}
         low, &\text{if}\: u_{CPU}^{PDU} \le cpu_{-}^{PDU} + \lambda \\
         medium, &\text{if}\: u_{CPU}^{PDU} \in [cpu_{-}^{PDU} + \lambda,\; cpu_{-}^{PDU} + 2\lambda)\\
         high,   &\text{if}\: u_{CPU}^{PDU} \ge cpu_{-}^{PDU} + 2\lambda. \\
\end{cases} \nonumber
\end{align}
PDU's minimum and maximum CPU usage, i.e., $cpu_{-}^{PDU}\equiv\min u_{CPU}^{PDU}$ and $cpu_{+}^{PDU}\equiv\max u_{CPU}^{PDU}$ are measured historically. Note that in the previous expressions we omit the reference to time for more clarity. The three segments are assumed to be of equal lengths, defined as $\lambda = \frac{cpu_{+}^{PDU} - cpu_{-}^{PDU}}{3}$. 
Piecewise linear models in the context of machine and cluster level power estimation were previously proposed in \cite{davis2011no}, where the Multivariate Adaptive Regression Splines (MARS) \cite{friedman1991multivariate} was used to automatically learn the regime switching points (called knots). The Per-PDU model is defined using the three equal segments corresponding to low, medium, and high usage regimes, while achieving robustness (e.g., monotonically increasing power with cpu usage) and a desirable accuracy for the use cases of interest.

For each PDU, linear models are trained for each usage regime to estimate the power consumption of a given PDU as
\begin{equation}\label{eqn:perPDU}
\Hat{u}_{POW}^{PDU}=\alpha^{r} + \beta^{r} u_{CPU}^{PDU}.
\end{equation}
The intercept and slope corresponding to each regime, i.e., $\alpha^{r}$, $\beta^{r}$, $r\in\{low, medium, high\}$ are computed so that the weighted sum of squared errors is minimized. The model parameters in (\ref{eqn:perPDU}) are constant for each day and updated daily. The Per-PDU models are trained and evaluated daily, using the most recent $7$ days of PDU's power and CPU usage data. The recurrent, daily, training is done to adapt the model in case of hardware changes (deployment/decommission of servers). 
The training instances are weighted based on their recency, i.e., higher weights are assigned to more recent measurements to ensure proper adaptation to systematic changes in the fleet (e.g., hardware upgrades and new deployments). In particular, the pipeline uses $\frac{1}{1+d}$ to weigh data instances from $d$ days ago.

The predictive performance of the Per-PDU model is evaluated daily using power/CPU usage data from the next day (Section \ref{sect:perf_eval}). It is observed that the Per-PDU model is able to adapt to sporadic changes in workload properties and hardware infrastructure in $\sim 2$ days. Outside of these transient regimes, it is observed that the coefficients of a Per-PDU model change very slowly in time. 
Note that the Per-PDU model implicitly incorporates effects of real-time scheduling and frequency scaling within a PDU. 

In addition to the efficiency-aware workload management, the Per-PDU model is used for monitoring and near-term predictions of power/energy usage and electricity-based carbon footprint across Google datacenters. 
Also, these models are particularly useful when power measurements are delayed or power meters are broken.

\subsubsection{Unified Machine and PDU Models}
\label{ssec:unifiedm}
In this section, we discuss Unified Machine and PDU models that are particularly useful for long-term infrastructure planning. For this purpose, it is required to predict the power consumption of future PDUs (example: after addition of some planned capacity) at a given resource usage operating point. The Per-PDU model is not suitable for addressing and evaluating the impact of a future deployment of servers on power consumption since the model is mostly applicable to predictions in short terms. Furthermore, Per-PDU models are not relevant for PDUs which have not been built yet. In order to effectively plan power for (existing or future) PDUs with changing server compositions, we need power models that can be generalized across PDUs and are able to incorporate knowledge about server deployment plans. This would enable PDU power prediction for long-term future planning scenarios.

The Unified Machine model sees PDUs as a collection of machines types, and the Unified PDU model sees PDUs as a collection of platform families. They can be trained with data from one PDU and predict power in another PDU (Section \ref{sect:perf_eval}).

Random Forest \cite{liaw2002classification} is used to model nonlinear relationships between the features (such as CPU usage/utilization, hardware characteristics, workload types) and the output, i.e., the power consumption of any machine (in case of Unified Machine model) or PDU power (in case of Unified PDU model) across Google's fleet. Random Forest is a powerful regression model and has been widely used to learn nonlinear relationships between variables in various research areas (e.g. \cite{mutanga2012high}). Specifically, Random Forest can handle high dimensional data, which may include both continuous and discrete variables. 
The abundance of power consumption and resource usage data for all machines and PDUs across Google's datacenter fleet makes this type of models a good candidate for predicting power consumption.

To train the Random Forest regressor, Sklearn \cite{sklearnwebsite} package is used with the default values for most of the hyperparameters, and the Mean Squared Error as the minimization objective. Hyperparameters such as number of trees, maximum tree depth and minimum number of samples in each leaf node were manually tuned so that the training and validation errors attain convergence with the selected number of samples and show no further improvement. We randomly selected one week of data for training. Selecting any different set of training data, as long as it contains data instances across all CPU usage levels results in almost identical predictive performance. 

\begin{figure*}
\centering
    \centering
    \includegraphics[width = 0.8\textwidth]{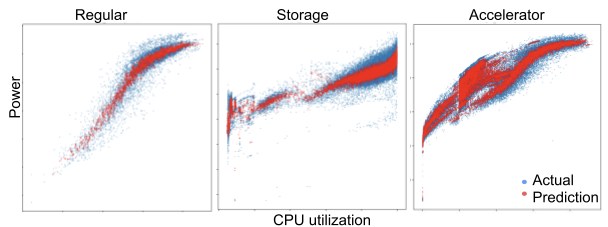}
    \caption{Nonlinear relationships between machine power and CPU utilization for regular, storage and accelerator machine types (in blue) and prediction of the Unified Machine model (in red).}
    \label{fig:machine_scatter}
\end{figure*}

\textbf{Unified Machine Model}
estimates power consumed by all machines deployed in a PDU, which are then added up to predict the PDU's power. To predict machine level power consumption, the industry has been mainly using the approach in \cite{fan2007power}, which interpolates power usage based on a straight line connecting machine's idle power (corresponding to $0$ CPU usage) and maximum power (corresponding to the maximum CPU usage). However, in reality, the relationship between machine CPU usage and power consumption is nonlinear as shown in Figure \ref{fig:machine_scatter}. The Unified Machine model is able to learn this nonlinear relationship between power and CPU usage (Figure \ref{fig:machine_scatter}) for each machine type that exist in the Google fleet. 

Most of the machines in datacenters can be categorized into three groups: typical compute machines, storage machines, and machines with accelerators (TPU or GPU). For all three categories of machines, CPU utilization can be an indicator of power usage, as shown in Figure \ref{fig:machine_scatter}. As a result, using CPU utilization as a feature can produce accurate results for all three types of machines. Moreover, as shown in Figure \ref{fig:machine_scatter}, the relationship between CPU usage and power usage can vary for different machine types. Thus, features related to the hardware type are also included.

The training data set used for the Unified Machine model includes the following features:
\begin{itemize}
\item $C^m$: incorporates a detailed description of machine's motherboard family, CPU type and size, memory quantity $\times$ size, SSD quantity $\times$ size, disk quantity $\times$ size, etc. (e.g., Xeon machine with $n_{CPU}$ CPUs, $n_{RAM}\times 32GB$ memory, and $n_{disk}\times 1TB$ disk). Each machine configuration has a unique code in the format of a 5 digit number. Machine configuration is a discrete feature, and is converted into a series of binary features using the one-hot encoding approach \cite{potdar2017comparative}. One-hot encoding results in creation of many more new features. We then use feature selection to remove the features that does not impact the model performance.
\item $P_-^m$, $P_+^m$: the idle and maximum power usage of a machine with a given configuration respectively.
\item $l^d$: indicates if a machine is shared by various applications (i.e., workloads), or is dedicated for a specific purpose (e.g., Google search). This feature is one-hot encoded.
\item $u_{CPU}^{m}$: machine's CPU usage.
\end{itemize}
The objective is to develop a mapping $f(.)$ between the above-described features and machine's power consumption, expressed as
$\hat{u}_{POW}^{m} = f(C^m,P_-^m,P_+^m, l^d, u_{CPU}^{m})$.

To reduce the computational complexity of the training phase, we randomly sample $30000$ 5-minute data instances across all machines with a specific machine configuration and dedicated label from a randomly selected week of data. 
To make sure that the model accurately predicts power across the entire CPU usage regime, stratified sampling \cite{neyman1992two} is used, i.e., machine's CPU utilization range is split into $10$ equally spaced buckets, after which the data is sampled randomly from each bucket. 

The total power consumed by the machines on the PDU $\Hat{u}_{POW}^{IT}$ is estimated by summing up its machines' power predictions, i.e.,
\begin{align}\label{eqn:umm}
\Hat{u}_{POW}^{IT} &= \sum_{m\in PDU} \Hat{u}_{POW}^{m}\nonumber \\
&= \sum_{m\in PDU} f(C^m,P_-^m,P_+^m, l^d, u_{CPU}^{m}).
\end{align}
Finally, the total PDU power prediction, $\Hat{u}_{POW}^{PDU}$, is computed by adding the estimate of the power usage of its networking and cooling equipment, $u_{POW}^{O} := u_{POW}^{PDU} - {u}_{POW}^{IT}$, to the total predicted machines' power $\Hat{u}_{POW}^{IT}$. Since $u_{POW}^{O}$ exhibits small variations across time, and is significantly smaller than PDU's total power consumption, it is estimated using its average measurement, $\bar{u}_{POW}^{O}$, from the previous day.

The high accuracy of the Unified Machine model (as discussed in Section \ref{sect:perf_eval}) can be attributed to three factors: 1) for each machine type, we have data from thousands of machines and in more diverse operation regimes, from 0 CPU usage to almost $100\%$ CPU usage. This gives us ability to predict PDU power accurately at any CPU usage operating point (aggregated CPU usage across all of its machines); 2) by aggregating power estimates for a large number (tens of thousands) of machines to obtain the PDU power prediction, the prediction error on individual machines cancels out, reducing the overall prediction error; and 3) the Unified Machine model predicts PDU power with more detailed system-related features, which leads to high prediction accuracy.

\textbf{Unified PDU Model} estimates PDU power consumption $u^{PDU}_{POW}$, using PDU level hardware and resource usage features as listed below:
\begin{itemize}
    \item $P_-^{PDU}$, $P_+^{PDU}$: sum of the idle and maximum powers across all machines in a PDU.
    \item $N_{F_i}$: total number of machines per platform family. For example, all computing machines with Intel CPU belong to the platform family named Intel. Another category of machines is storage which, depending on the type of storage (e.g., SSD), has a few different platform families. Overall, there are $10$ platform families across Google's datacenter fleet, say $F_1,...,F_{10}$, with the corresponding number of machines within a given PDU denoted as $N_{F_1},...,N_{F_{10}}$, which are used as features. 
    \item $u_{CPU}^{F_i}$: sum of CPU usage of all machines per platform family in a PDU, that is, $u_{CPU}^{F_i}:= \sum_{m \in F_i} u_{CPU}^m, i = 1,...,10$, are used as the resource usage features. 
    Note that each platform family contains several machine configurations.
    \item $P^N$: Power drawn by the networking equipment within a PDU is not metered separately, and we use its maximum value as a proxy. A small dynamic range of network-related power consumption justifies the approximation (also previously observed in \cite{dayarathna2015data, fan2007power}). 
    \item $P^C$: Similar to $P^N$, maximum power drawn by the cooling equipment within a PDU is used as a proxy for its power.  
    \item $A^P$: A categorical feature is used to identify the type of each PDU’s power architecture from the three available types, and is one-hot encoded.
\end{itemize}
The objective is to develop a mapping $g(.)$ between the above described features and PDU power consumption, expressed as
\begin{align}\label{eqn:upm}
    &\hat{u}_{POW}^{PDU} = \nonumber \\
    &\quad g(P_-^{PDU}, P_+^{PDU},\{N_{F_i}\}_{i=1}^{10},\{u_{CPU}^{F_i}\}_{i=1}^{10},P^N,P^C, A^P).
\end{align} 
The model is trained using ~$1$ million instances of PDU level CPU and power usage measurements at $5$-minute granularity, and across all PDUs within Google's fleet. 

Both Unified models (Machine and PDU) are trained using the same week of data to provide a consistent baseline for models comparison. The prediction accuracy of the Unified PDU model significantly depends on the distribution of the platform family mix and CPU usage regimes captured by the training data, as discussed in more detail in Section \ref{sec:perf_eval_power_drop}.
\subsection{Models' Complexity}

The Per-PDU model is essentially a piecewise linear regression model, the complexity of which is $O(n)$ with $n$ being the sample size of the training data \cite{9317826}. The time complexity of the Random Forest models is $O(n\log n)$, as discussed in \cite{9317826}.

The Per-PDU models are decoupled and can be trained separately for each PDU, and thereby, completed within minutes.
The Random Forest model also has parallelization built in and can be finished in around 2 hours. In addition, our validation results have shown that the Unified models are relatively stable over time, thus not requiring to be retrained often, as is typical with offline machine learning applications.

The space complexity of machine learning models are closely related to the number of parameters in the models. Our models typically do not take up a lot of storage space due to relatively small number of hyperparameters compared to the dataset being used in training and evaluation phases. Thus space complexity is not a constraining factor. The models presented in this paper are deployed within Google's production system to predict PDU power, which demonstrates scalability of the models at high-dimentional industrial levels.

\section{Performance evaluation through targeted use cases}\label{sect:perf_eval}
The previously discussed power models are rigorously validated for all PDUs across Google's datacenter fleet. The predictive performance of the models is analyzed while taking into consideration the use cases of interest: 
\begin{enumerate}
\item near-term power estimation and, consequently, carbon accounting (\cite{energy2019data, greenieee}); 
\item near-term load shaping, which includes peak power shaving (\cite{fan2007power}), load drop experiments (e.g., grid level demand response \cite{LiuChenBashWierman2012}), as well as carbon- and cost-aware load shifting (\cite{liu2015datacenter, radovanovic2020our, energy2019data}); 
\item long-term planning and rightsizing of future datacenter infrastructure, including upgrades and deployments.
\end{enumerate}

For the majority of the use cases listed above, the main goal is to accurately predict PDU power consumption as a function of its CPU usage (or utilization), hardware and workload properties. There are different performance metrics that could be used to quantify prediction's deviation from the actual power measurements. The metrics should provide a reasonable comparison across different PDUs regardless of their maximum power capacities. To that end, Mean Absolute Percent Error (MAPE) is chosen to evaluate performance of the power models, and is defined as
\begin{equation}
    MAPE^{PDU}(d) = \frac{100}{|\mathcal{T}|}\sum_{t}\frac{|\hat{u}_{POW}^{PDU}(t)-u_{POW}^{PDU}(t)|}{u_{POW}^{PDU}(t)} \label{eq:mape},
\end{equation}
It should be noted that all the performance results in this section are computed using data instances outside of the training data set.

\subsection{Use case \#1: near-term power estimation}
To test $5$-minute power prediction accuracy across all PDUs fleetwide, the proposed models are trained with seven days of historical data, and their performance (as defined in (\ref{eq:mape})) is evaluated using the actual PDU resource and power usage, and other collected features from the next day.

\begin{figure}
    \centering
    \includegraphics[width=0.8\linewidth]{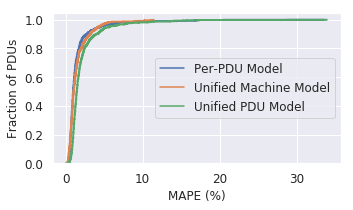}
    \caption{Fraction of the total number of PDUs with average daily MAPE less than a given value.}
    \label{fig:mape_all}
\end{figure}

To compare the performance across all $3$ proposed models, a $2$-step evaluation analysis is conducted: 

1) For a randomly selected week, each PDU and model, average daily MAPE is computed as 
\begin{equation}
\overline{MAPE}^{PDU}:=\frac{1}{7}\underset{d\in week}{\sum}MAPE^{PDU}(d).
\end{equation}
Then, as shown in Figure \ref{fig:mape_all}, we capture the fraction of the total number of PDUs with the average daily MAPE smaller than a given percent value. As can be observed in Figure \ref{fig:mape_all}, all three models have similar performance, with MAPE smaller than $5$\% for more than $90$\% PDUs, and MAPE smaller than $10$\% for more than $99$\% of PDUs. The uniformly low MAPE across all models makes them good candidates for short-term estimation and forecasting of PDU average power, which is then aggregated to obtain clusters' and campuses' power predictions. 

\begin{figure}
    \centering
    \includegraphics[width = 0.8\linewidth]{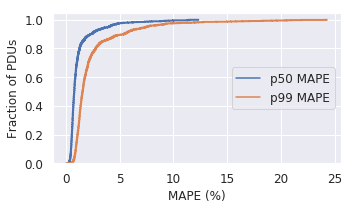}
    \caption{Fraction of the total number of PDUs with the median and $99$th percentile of daily Per-PDU's MAPEs less than a given value.}
    \label{fig:mape_long}
\end{figure}

2) To test the uniformity of the evaluated performance across time, daily MAPEs are calculated for each PDU across a year-long time horizon, i.e., $\{MAPE^{PDU}(d)\}_{d\in year}$. Both the Unified PDU and Unified Machine models are inherently time invariant since their training data covers the full range of values of the features. On the other hand, to evaluate the temporal insensitivity of the Per-PDU model, $50$th and $99$th percentiles of $\{MAPE^{PDU}(d)\}_{d\in year}$ are computed for each PDU. We then compute the fraction of PDUs with their daily MAPE median ($50$th percentile) and $99$th percentile smaller than any given percentage value. As shown in Figure \ref{fig:mape_long}, the Per-PDU model's overall performance exhibits some performance deviations across time.


\subsection{Use case \#2: load shaping}\label{sec:load_shaping}

While the proposed models exhibit similar performances in MAPE, this is not any more the case if the goal is to estimate PDU power when there is a drop or changes in datacenter compute load in response to:
\begin{enumerate}
    \item datacenter safety or maintenance events,
    \item grid level demand response events,
    \item intraday datacenter load optimization for its daily electricity-based carbon footprint.
\end{enumerate}
The Per-PDU model's piecewise linearity provides a suitable framework to evaluate the impact of changes in CPU usage on power usage, which can then be easily integrated into cost and carbon-aware optimizations designed to shape compute load by shifting computing tasks in time and space, across datacenter clusters (\cite{LiuChenBashWierman2012}, \cite{liu2015datacenter}, \cite{radovanovic2020our}).

Next, the proposed models are evaluated based on their ability to predict the variation in power consumption caused by significant changes in compute load. Then, we discuss the methodology and importance of accurate power modeling and its error characterization for controlling PDU peak power consumption.

\subsubsection{Power drop experiments}\label{sec:perf_eval_power_drop}

\begin{figure}[h]
    \centering
    \includegraphics[width=0.8\linewidth]{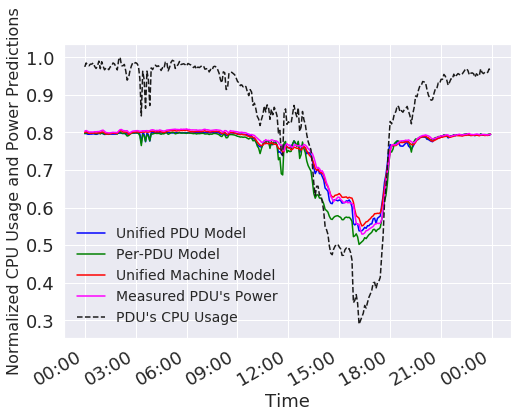}
    \caption{Example of a significant load drop in one of the experimental PDUs.}
    \label{fig:pd_experiment}
\end{figure}

There are several applications in which a model's capability to accurately extrapolate power usage estimation outside the previously seen regimes of compute load is critical. Examples of such applications are: 1) risk-aware planning of large workload migration across different clusters; and 2) significant, planned drops in compute load to respond to grid level demands. To evaluate the applicability of the proposed approaches to such scenarios, power drop experiments were conducted in $8$ clusters spanning all types of power architectures across Google's datacenter fleet. The planned usage drop was performed between 3:50 pm and 5:25 pm on a given day by progressively terminating non-critical tasks based on their priority (starting with the lowest priority). Performance of the proposed models was evaluated within the testing interval. Figure \ref{fig:pd_experiment} shows an example of the actual power and CPU usage of an experimental PDU, along with the corresponding predictions of the three models.
\begin{figure}
    \centering
    \includegraphics[width=0.8\linewidth]{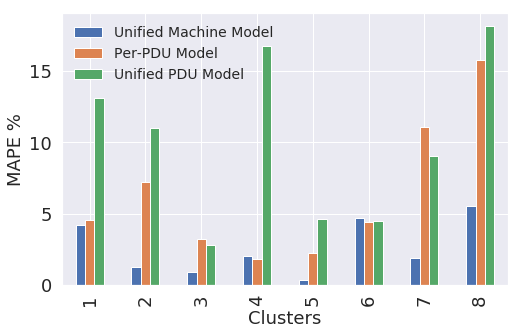}
    \caption{Cluster level performance of the proposed models for eight experimental clusters during the planned power drop event.}
    \label{fig:mape_exp}
\end{figure}

To evaluate performance of the power models, average MAPE is computed across all PDUs within a given cluster and using the data instances within the testing interval, i.e., $\frac{\sum_{PDU\in \mathcal{C}}MAPE^{PDU}}{|\mathcal{C}|}$. The experimental results in Figure \ref{fig:mape_exp} demonstrate that the Unified Machine model has the best predictive performance during the power drop events, with an average MAPE of less than $6$\% across all the test clusters. 
However, there are cases where the Per-PDU and Unified PDU models perform comparably well. Clusters $3$ and $6$ in Figure \ref{fig:pd_experiment} are two examples where both the Per-PDU and Unified PDU models generate accurate predictions. This is not surprising given that the dropped, experimental, CPU usage is captured in the training data. There are, however, scenarios, where the Per-PDU model's MAPE is less than 5\% even when the low CPU usage regime is not present in its training data (e.g., see clusters $1$, $4$ and $5$ in Figure \ref{fig:pd_experiment}), which implies that the extrapolated, low-usage-regime segment of the Per-PDU model could still be an effective predictor of PDU power.

\subsubsection{Power capping}
As discussed in the Introduction, another benefit of accurate power models is software-controlled power capping, used to limit maximum power usage of a PDU by controlling its CPU usage. To achieve this goal in a risk-aware manner, characterizing model's underprediction error (i.e., when predictions are lower than the actual power measurements) is critical to ensure that the target power usage threshold of a PDU is respected by properly setting its CPU usage limits. To that end, the worst (largest) $90$-day underprediction percent error is computed daily across time instances that correspond to PDU's high CPU utilization (and, therefore, high power utilization) regime:
\begin{equation}
    WUPE^{PDU}(d) =100 \cdot \underset{t\in [d-1, d-90]} {\underset{u_{CPU}^{PDU}(t)\in high}{\max}} \frac{(u_{POW}^{PDU}(t)-\hat{u}_{POW}^{PDU}(t))}{u_{POW}^{PDU}(t)}.
    \label{eq:worst_mape}
\end{equation}

To avoid tripping a PDU's circuit breakers, the power capping system typically responds by throttling low priority compute tasks, which reduces CPU (and, thereby, power) usage of the PDU. Therefore, statistical properties of the underprediction error of the models can be used to define user-facing service level objectives (SLOs) of the power capping system. SLOs are designed to set user-perceived expectations during unexpected workload spikes and therefore, inevitable power capping events.

\begin{figure}
    \centering
    \includegraphics[width=0.8\linewidth]{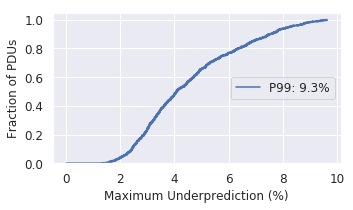}
    \caption{CDF of $\max_{d\in {5\;\text{month horizon}}} WUPE^{PDU}(d)$ across all PDUs in the datacenter fleet.}
    \label{fig:under_prediction_cdf}
\end{figure}

For each PDU, maximum $WUPE^{PDU}(.)$ of a Per-PDU model is computed over a $5$-month time horizon, i.e., $\max_{d\in {5\;\text{month horizon}}} WUPE^{PDU}(d)$, the CDF of which is shown in Figure \ref{fig:under_prediction_cdf}. The maximum of the worst underprediction errors across all PDUs are then used to obtain the fraction of the total number of PDUs with the corresponding maxima less than a given value. The analysis shows that for $99$\% of PDUs, the worst underprediction error is less than $9.3$\%. It is observed that larger underprediction errors in some PDUs typically happen due to the unpredicted machine upgrades, to which the Per-PDU model typically adjusts within 2-3 days.

\subsection{Use case \#3: rightsizing infrastructure upgrades}

The structure and accuracy of the Unified Machine model make it suitable for cost- and performance-aware infrastructure planning, commonly referred to as rightsizing. Typical planning scenarios require: 1) provisioning power consumption after deployments/decommissioning of machines, 2) analysis of the long-term impact of platform mix in a power domain on its power consumption, and 3) studying the effect of large workload migrations across a datacenter fleet on its power consumption. Such scenarios require long-term forecasting (monthly, quarterly, etc.), where PDU power and resource usage data are either unavailable, or the predicted operating regime is different from the historically observed pattern. Among the proposed models, the inherent nature of the Unified Machine model makes it an ideal candidate for mapping long-term compute infrastructure plans and workload trends to power consumption forecasts. The Unified Machine model estimates power consumption of a PDU by aggregating its machine level predictions, where machine level training instances capture the full dynamic range of CPU utilization (see Subsection \ref{ssec:unifiedm} for more details). This enables the model to predict power consumption at various operating regimes with higher accuracy.

\begin{figure}
    \centering
    \includegraphics[width=0.8\linewidth]{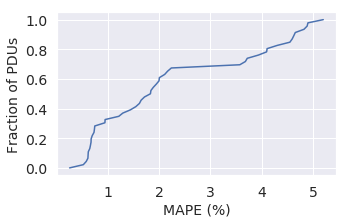}
    \caption{Fraction of the hold-out PDUs with $MAPE^{PDU}$ smaller than a given percent error.}
    \label{fig:hold_out_cdf}
\end{figure}

To evaluate performance of the Unified Machine model in predicting power consumption of new (future) PDUs, $47$ PDUs across all power architectures are randomly selected and removed from the training dataset. The Unified Machine model is then retrained on the reduced training dataset, and used to predict PDU level power for each of the hold-out PDUs at $5$-minute time granularity. The model is tested on a week of data outside the time period used in the training dataset. For each hold-out PDU, $MAPE^{PDU}$ is computed using all data instances within the test week/dataset. When averaged across all tested PDUs, the computed MAPE is $2.23$\%, while PDU level MAPEs are below 5\% for the majority of the tested domains (Figure \ref{fig:hold_out_cdf}). This is consistent with the prediction accuracy for PDUs included in the training data set, and indicates the ability of the Unified Machine model to provision power of unseen PDU configurations.

\section{Conclusions}\label{sect:conclusion}
In this paper, design of two types of statistical power models is discussed along with rigorous validations for accuracy, simplicity, interpretability, and applicability to all hardware configurations and workloads across hyperscale datacenters of Google fleet. The two types of models (already deployed in production) are based on: 1) individual machine level data, and 2) the entire MW-scale datacenter PDU level.  The  models  are  developed to accommodate several use cases, including: cost- and carbon-aware load management, power provisioning, and infrastructure rightsizing. To the best of our knowledge, this is the largest scale datacenter power modeling effort based on real operational data, in both the scope of diverse use cases  as well as the variety of hardware configurations and workload types  used for modeling and validation. It was demonstrated that the proposed statistical modeling techniques, while simple and scalable, can predict power with less than 5\% Mean Absolute Percent Error (MAPE) for more than 95\% diverse PDUs. This performance matches the reported accuracy of the previous state-of-the-art  methods,  while  using  significantly  less  features and  covering  a  wider  range  of  use  cases.

The models facilitate several decision-making scenarios that datacenter planners and compute providers face today, which will be a focus of our future efforts. This will include enhancing and using the models for reducing carbon footprint and reducing energy costs, increasing utilization of existing infrastructure, and rightsizing future datacenter designs. Moreover, the developed models will be used for building intelligent computing systems capable of shifting computing tasks in time and space in order to reduce datacenter power peaks and electricity-based carbon footprint \cite{greenieee, radovanovic2020our}.
\bibliographystyle{IEEEtran}
\bibliography{IEEEabrv,reference.bib}
%

%








\end{document}